\documentclass{article}

\usepackage[english]{babel}

\usepackage[letterpaper,top=2cm,bottom=2cm,left=3cm,right=3cm,marginparwidth=1.75cm]{geometry}

\usepackage{amsmath}
\usepackage{amssymb}
\usepackage{graphicx}
\usepackage[colorlinks=true, allcolors=blue]{hyperref}
\usepackage{caption}
\usepackage{subcaption}
\usepackage{float}
\usepackage[numbers,authoryear]{natbib} % Added for author-year citation style
\usepackage{authblk} % For multiple authors and affiliations
\usepackage{hyperref}
\usepackage{setspace}
\usepackage{enumitem}

\providecommand{\keywords}[1]
{
  \small	
  \textbf{\textit{Keywords---}} #1
}

\newcommand{\DKL}{D_{\text{KL}}}

\title{The Physics and Metaphysics of Social Powers: Bridging Cognitive Processing and Social Dynamics, a New Perspective on Power through Active Inference
}

\author[1,2,3]{Mahault Albarracin}
\author[4,5]{Sonia de Jager} 
\author[6]{David Hyland}
\author[7]{Sarah Grace Manski}

\affil[1]{VERSES.ai, Los Angeles, California, USA}
\affil[2]{Université du Québec à Montréal, Département d'informatique, Montréal, Québec, Canada}
\affil[3]{Université du Québec à Montréal, Institut de Sante et Societe, Montréal, Québec, Canada}
\affil[4]{Erasmus School of Philosophy, Erasmus University, Rotterdam, Netherlands}
\affil[5]{Noise Research Union (NRU)}
\affil[6]{Department of Computer Science, University of Oxford, Oxford, UK}
\affil[7]{Spatial Web Foundation, Los Angeles, California, USA}

\date{} % Leave blank for no date, or specify a date

\begin{document}
\maketitle

% \section{Conflict of Interest}
% There is not conflict of interest to note for this work.

% \newpage

\begin{abstract}
    The concept of power can be explored at several scales: from physical action and process effectuation, all the way to complex social dynamics. A spectrum-wide analysis of power requires attention to the fundamental principles that constrain these processes. In the social realm, the acquisition and maintenance of power is intertwined with both social interactions and cognitive processing capacity: socially-facilitated empowerment grants agents more information-processing capacities and opportunities, either by relying on others to bring about desired policies or ultimately outcomes, and/or by enjoying more information-processing possibilities as a result of relying on others for the reproduction of (material) tasks. The effects of social empowerment thus imply an increased ability to harness computation toward desired ends, thereby augmenting the evolution of a specific state space. 

Empowered individuals attract the attention of others, who contribute to increasing the scale of their access to various policies effectuating these state spaces. The presented argument posits that social power, in the context of active inference, is a function of several variables, among them: an individual’s or group’s ability to attract the attention of others with the ability to facilitate desired policies, and an individual’s or group’s capacity to effectively process information. As a result of its power-amplifying effects, this extended computational ability also buffers against possible vulnerabilities. Semantic social artifacts (narratives, ideologies, histories, representation, etc.) as attentional scripts (ensuring the coordination of social patterns: psychological schemata, habits, groupthink, insulation, policies, etc.:  \citep{albarracin2021variational}), are the structures under and by which agents mediate and leverage power, and are thus this article’s main objects of analysis.

Narratives are social mechanisms which have the ability to make or break power \citep{blakeley2024vulture}. Not-yet-engaged scripts (speculations, fictions, etc.: future forecasting) can be understood as semantic attractors enabling—relatively—risk-free learning \citep{dejager2022fiction}.  Possibilistic cognitive explorations in the form of, e.g., (fictional) narratives allow agents to consider counterfactual scenarios and minimize uncertainty without exposure to “real-world” consequences. This mechanism frames both the evolutionary advantage and persistent appeal of narrative engagement in human culture. In the context of power, leveraging “strong narratives”—such as “us versus them” reductionisms—greatly minimizes uncertainty (technically: variational free energy, the divergence between what is expected and what is actually experienced) about an agent’s or group’s current predicaments, and thus become appealing (sometimes self-fulfilling) policies that bring about their effects, resulting in collective behaviors which can be beneficial or detrimental to the greater whole \citep{albarracin2022epistemic}.

The effects of variational free energy-minimization can be observed at the level of the basic structures of language: (semantic) attractors such as redundancy, rhyme and alliteration, can be understood as patterns which aid memory, communication and information-retrieval policies: they permanently reappear, pooling attention to the “same” phenomena, thus greatly reducing the need to decode and recode linguistic interaction anew. By way of the same uncertainty-minimizing mechanisms, (simple or complex) narrative scripts create cognitive situations where agents can reduce complex aspects of the world down to preferred and/or already-identified scenarios. Unlike “direct” experience, where prediction errors present immediate risks, narratives provide speculative semantic spaces for exploring models against hypothetical outcomes. With regard to social power, the problems immediately emerge in situations where collective attention controlled by agents already benefiting from dominant narratives can lead to, e.g., the exacerbation of misinformation, echo chambers, conspiracy theories, etc. (ibid.), in an attempt to retain attention and therefore power. In political communication and organization, where the proliferation of scripts can be understood as an optimization of the learning/risk trade-off inherent in active inference, this is of crucial attention, particularly at a time when the democratic, technological and futurological organization of life is in question.

We propose that individuals wield power not only by associating with others possessing desirable policies, but also by enhancing their ability to intake and compute information effectively. This dual mechanism is argued to create a cyclical, reinforcing pattern wherein the empowered are able to incrementally expand the scope of policies and state spaces available to them while minimizing risk-exposure. The theoretical framework developed here has implications for understanding both the cognitive foundations of script engagement in the mediation and evolution of power, as well as strategies for their possible modulations. Formalizing and modeling these mechanisms demands special attention in the context of futurological prospects. This article reorients what has been previously framed in terms of territorial, material and cultural power towards what can now be understood in terms of informational and computational power. All of these have received attention as forms of capital; a term we can reframe under active inference as possibilistic power, i.e., the creation of the future.
\end{abstract}


\section*{Introduction: The Nature of Power
}

The concept of \textit{power} has a long history, and can be explored at several scales. In this article, our interest is to explore the concept as a phenomenon that can be understood as unfolding across scales, by framing and formalizing it through active inference. In the context of social organization dynamics, one of the most compelling elucidations of what constitutes the nature of power was presented by Michel Foucault:

\begin{quote}
``... power must be understood in the first instance as the \textit{multiplicity of force relations immanent in the sphere in which they operate and which constitute their own organization}; as the process which, through ceaseless struggles and confrontations, transforms, strengthens, or reverses them; as the support which these force relations find in one another, thus \textit{forming a chain or a system}, or on the contrary, the \textit{disjunctions and contradictions which isolate them from one another}; and lastly, as the \textit{strategies} in which they take effect...'' (\citep{foucault1990history}, pp. 92–93, our emphasis in bold).
\end{quote}

The phenomenon, as well as the representations of what precisely constitutes the social acquisition and maintenance of power, is thus a complex series of multidirectional processes which cannot be simplistically or hierarchically schematized, but must be analyzed in terms of interactions between agents and the constraint-regimes they are embedded within. Traditionally, the framing of socially-facilitated empowerment—such as that of the state, or of a large corporation—has been understood as the effectuation of (often top-down) strategies, tactics, policies, narratives (i.e., \textit{scripts}, as will be argued later) which grant empowered agents/systems more \textit{leverage}, that is: more influence on future states \footnote{ Note: here we refer to the meaning of state as “condition”, not in the political sense. However, the double entendre is opportune in this context.
}. We will frame this in terms of information-processing capacities and opportunities, where powerful agents either rely on others to bring about desired policies (i.e., exploitation: in the biological and sociological senses), or enjoy more information-processing possibilities as a result of relying on others for the reproduction of (material) tasks (i.e., exploration). The effects of social empowerment therefore imply an increased ability to compute information, thereby augmenting the evolution of a specific state space. Basing our arguments in active inference, as will be explained, will allow for an analysis of “[p]ower [a]s the probability that one actor within a social relationship will be in a position to carry out [their] will despite resistance,”(\cite{weber2009theory}, p. 152) in terms of \textit{actual} \footnote{Weber’s quote ends in “... regardless of the basis on which this probability rests.” (ibid.). We will challenge this through a material-informatic analysis, as the basis of this probability is not trivial. See, e.g.,  \citep{ramstead2023bayesian}, on the introduction of a “formal language for modelling the constraints, forces, potentials, and other quantities determining the dynamics of such systems, especially as they entail dynamics on a space of beliefs (i.e., on a statistical manifold)” for more background on the possible bridges between material and informatics in terms of Bayesian mechanics. Also note that the translation of “probability” in Weber has also been that of “ability”, “capacity”, and other terms, we take one of the most popular formulations here which also aids our purpose of framing this in terms of a probabilistic mechanics. 
} probabilities. We will present novel understandings of concepts such as scripts and attention, by analyzing power in terms of the physics of beliefs \citep{ramstead2023bayesian}, and the metaphysics of social organization, where metaphysics is understood as the possibilistic exploration of state spaces (i.e., the future).

The outline of the article is as follows: in the subsections that follow, we present a necessarily broad and brief elucidation of different understandings of power, primarily by way of philosophy and sociology, in order to set the focus on the concept of power as the control of attention via explorative and exploitative scripts, which can exist as shared metacognitive computing conditions (i.e., agents are relatively aware of the implications of scripts), coercive domination (where agents are subsumed under a narrative which does not benefit them), or: often an inevitable combination of both. In section 2, we explain the formal context of Active Inference as the theoretical ground which will sustain our analysis. In section 3, we synthesize the previous sections towards an explanation of social scripts and power, and how scripts can frame the concept of narratives, attention, coercion and resistance. In section 4, we look at power through the lens of Active Inference, and in section 5, we treat the formal framework integration. We conclude by presenting attentional strategies which can result in alternative modulations of power. 

 \subsection{Power-pooling in, through and between agents}

Gravitational metaphors abound in language  \citep{shepard1987toward,lakoff2003metaphors,dennett1992self,dejager2021inevitably,kent2024mental}\footnote{ Lachlan Kent has further argued that the vestibular system, which regulates graviception, also affects the higher-order consciousness of selfhood \citep{kent2024mental}: gravity constrains physical proprioception and action, as well as the very sense of a narrative self \citep{dennett1992self}.
}). Such gravitational metaphors find analogues in the network sciences, whereby preferential attachment processes lead to additional power accruing increasingly to those already in possession of it \citep{barabasi1999emergence}. The idea of power being pooled towards complex social attractors was famously presented by Pierre Bordieu, who linked various concepts of capital (cultural, political, etc.) to power as social (at)traction, borrowing the metaphor from physical field theory\footnote{Just as electromagnetic fields exert forces on particles depending on their position and properties, Bourdieu conceptualized social fields as spaces where agents’ positions are (pre)determined by their access to different forms of capital (economic, cultural, symbolic, etc.). Bourdieu’s use of field theory therefore frames system structure and the dynamics of agency: a given field structures possible actions by defining what counts as accepted or legitimate, and habitus as inclinations; dispositions, determine how fields are navigated: possibilities. The ensuing social practices herein are the result of heuristics comparable to satisficing/fast and frugal heuristics, which in our context are formalizable through AIF.}. We expand on his legacy which takes the ontoepistemological and the empirical to be inseparable—and his desire to ground sociological analyses of habitus in a scientifically tractable manner—by framing these attentional attractors that move systems toward specific configurations, through Active Inference (AIF). 
Another key thinker of power in the 20th century was Hannah Arendt. Arendt’s relational idea of collective power frames its essence as communicative, public and permanent potential: people may gain momentum together, but power is never guaranteed. Coercion, (symbolic) violence and oppressive authority, in her framing, do not constitute power but are rather reductions of its possibility. In our context, this differentiation would imply the first presentation being shared cultural narratives towards explorative change, and the second as top-down exploitative practices which reduce the possibility of novelty in given structures (systems such as communities; identities, histories, etc.). Similarly, more recently, Grace Blakeley has analyzed the disempowerment of collective political imagination by presenting this as the \textit{disillusion}\footnote{ In our terms, perhaps, understood as the loss of attentional momentum.} with shared narratives: democratic engagement in the context of, e.g., British politics, has been affected by a sense of lack of agency (from “there is no alternative” all the way to the not-so “free” market\footnote{“This sense of unfreedom is grounded in the deep disparities of power that exist within capitalist societies—many of which are completely invisible. Most people are denied autonomy over their lives, yet we are consistently told that we are free to choose how we live. Life under capitalism means life under a system in which decisions about how we work, how we live, and what we buy have already been made by someone else. Life under capitalism means living in a planned economy, while being told that you are free.” (\citep{blakeley2024vulture}, p. 6, our emphasis in bold).}). This resulting political disenchantment was framed earlier by Mark Fisher in terms of the privatization or individualization of systemic stress \citep{fisher2009capitalist}: through strategic narratives which set public attention on individual responsibility, publics become overwhelmed by having to resolve, at the individual level, stress which is in fact systemic. 
Whether we are talking about a metastabilizing “invisible hand” or modeling swarm-organization \citep{wareham2023swarm}: we know of many systems that have infelicitous critical points when scaling: requisite variety demands that control systems maintain sufficient internal complexity (entropy) to match the complexity of what they aim to control. This can be understood at a number of levels (in modeling: computational tractability problems regarding communication between agents that have to do with temporal tracking and issues such as latency, for example), but at the level of social dynamics and the (in)efficiencies of power asymmetries, this is most obvious in the context of the power of attention: the assertion of a specific narrative, if it gains traction as a collective belief, will result in the simplification (and mobilization) of desire. That is: a strong-enough narrative can become a script which agents incorporate as part of their generative model, rather than as a meta-consideration (a highly possibilistic variable), existing on a dialogical landscape. In crude terms: the difference between the first and the second can be exemplified by the situation where an agent believes that any single solution can solve national scale problems, such as the deportation of immigrants supposedly solving every socio-economic issue in a very large nation.
Our framing allows for a new take on the dialectics between the individual and the collective. In terms of questions of information-processing capacity: not only are resource-distribution and the division of labor a systemic fact, but also: since the modulation and projection of future possibilities cannot be computed at the level of the individual due to its high complexity, this is how and why the minimization of uncertainty for social systems is, in fact, social. Agents seek alliances and policies which ensure cognition (i.e., information processing and future-projection) can be distributed \citep{vasil2020world}, as this allows for efficient practical traction because of “epistemic confidence in knowing, interpreting, and acting together” \citep{shteynberg2023theory}. This is the basis for any division of labor, where specialized roles and collaborative synergy optimize task execution. As demonstrated by \citep{ispolatov2012division}, division of labor in biological systems improves fitness by compartmentalizing incompatible processes, while \citep{labella2007division} shows its application in robotics, enhancing adaptability and resource efficiency. \citep{marzband2017distributed} show that coalition-based strategies in distributed energy resources maximize collective benefits, offering a model for scalable and cooperative systems in human and artificial contexts. This is in contrast with individualized agency—which is a prerequisite for social coercion—where one’s estimation of “divergence in mental states between self and other(s)” allows for the “prediction and interpretation of others’ behavior” \citep{shteynberg2023theory}, resulting in shorter-range practical effects. 
Remaining politically-agnostic in any analysis presents us with some questions: what can we assume to actually know about each other when we act collectively? The concept of a ‘self’ as a single agent seems particularly questionable, especially as all selves exist within a regime of constraints which exert effects on their supposed autonomy. A famous response to the Foucauldian take on power and the individual came from Gilles Deleuze. His concept of the “-dividual” replaces the image of, e.g., the factory worker as a subject of disciplinary capitalism, with the image of a distributed subjectivity, e.g., the owner of a credit card as a subject of libertarian debt, occupying various spatiotemporal scales and modulated not by clock-time or physical enclosure, but by probabilities and contractual fixation/limitation \citep{deleuze1992postscript}. The latter condition emerges, Deleuze shows, as a result of novel paradigms of technocapitalism which present new constraint regimes. Distribution, collectivity, or shared sameness, therefore, do not always equal autonomous collective traction, but can also be the result of policies which pool power towards attractors, where the agents implicated within them may not even be aware of their implication (the examples in capitalism are vast: we have very little idea of the complexity behind the products we buy into, but we are coerced into buying nonetheless\footnote{For more on this theme, see “Post-control Script-societies” \citep{dejager2023semantic}.}). In the interest of this article, the concept tying all these processes towards an analysis of power dynamics is that of attention. We will briefly treat this below (and in later sections) before moving to section 2 for an exposition of the AIF framework.

\subsection{Attention as power: shared narratives and awareness or coercion}
Exerting social control over others’ actions begins at the level of attention: whether this be by attracting attention (“follow me!”), or directing attention (“look at that!”). Our interest in formalizing certain aspects of power dynamics is directing attention towards (metacognitive and collective) scripts with the capacity of empowering responses to counter dominant strategies which, currently, minimize uncertainty for the very few\footnote{Again, often by allocating too much importance to the atomized individual, which is a predictably convenient way to simplify the actions of complex distributed processes such as individuals, into overseeable patterns. There is nothing essentially “wrong” with this, this is the way perception tracks reality. However, revealing these dynamics allows us an entrance into the ways in which we may modulate (the effects of) these phenomena/concepts.}. Shared structures of meaning can exist as collective metacognitive computing conditions, i.e., conditions in which agents are—relatively—aware of the implications of the scripts they follow, or coercive domination: where agents are subsumed under a script which, often, does not benefit their uncertainty-minimization on the long run. If we understand successful narratives as those which exert the most effect\footnote{Often “fast and frugal” \citep{gigerenzer2008heuristics,simon1978rationality} heuristics and narratives, which are low-risk and, in most political contexts, imply promises or solutions impossible to accomplish: “you have been lied to”, “the other is the problem”, etc.}, analyzing them in terms of attention seems logical: in the context of uncertainty-minimization, simple narratives have the capacity to dominate vast regimes of attention (economies), as they are scripts which minimize the complexities of life into low-computation, predictable visions\footnote{Whether these are dialogical and representational/symbolic or surreptitious and encrypted: we’re dealing with different strategies, but similar results.}. 

In the transition between an economy centered around material and spatial capital, towards one of temporal, speculative capital, we can straightforwardly connect attention to computation by understanding the implicit link between risk-assessment and investment: (possible) future gains ought to be calculated, and directing investment begins with attentional gains\footnote{“A sufficiently large financial institution has the power to direct investment into certain technologies and therefore to determine which futures are available and which ones are foreclosed. And all these firms can consolidate their power by buying up competitors, erecting barriers to entry, and crushing workers’ attempts to organize, further insulating them from competitive pressure and giving them significant authority both within their domains and over society as a whole.” \citep{blakeley2024vulture}.}. Beyond the strictly financial realm, in the context of futurology and planetary survival, the concept of AI technologies as “powerful” grounds the arguments surrounding possible existential risks. E.g., Cappelen et al. rely on the concept of “power” in order to assess to what extent humans may or may not lose control over their own technological creations \citep{cappelen2024ai}. Likewise, other authors make claims for “powerful yet narrow AI systems” as incapable of accommodating the semantic complexity which grounds mortal life \citep{ororbia2023mortal}\footnote{They continue: “One might view the future as residing in not viewing intelligence as a difference engine but instead as an artificial form of sentience: one that is capable of self-healing and self-repairing with autonomy — its existential imperative being to persist in (generalized) synchrony with its world.” (ibid.). The questions of “autonomy” and “synchrony” are precisely the ones we are after in this article, by observing how power, in mortally-determined social systems, is driven by attention.}. What these assessments have in common is a concern with attention as a resource: a collectively predicted future is one where attention is directed towards synchronized goals (whether this be investment, alignment, etc.).

Ultimately, because potential power (as attention) is both a material and an informational force, “seizing the means of computation” (Doctorow, 2023) follows in line with a Marxist analysis of capital in our moving paradigm: from materialist analyses of territorial and labor time domination, through a semantic-sociocultural analysis \citep{bourdieu2018structures}, to informational analyses where we ought to think about the production of the future by means of projective computation. Power is thus based on the controlling of attention by way of scripts (again: narratives, representation, identities, symbolisms, etc.) and its current substrate in our contemporary landscape is computation: both in the flesh and in silico. As we will see in section 3, power’s dual nature (as force attraction and information computation) in the social realm thus implies the use of scripts as they are the means by which agents gain or lose the ability to direct the attention of other agents (with access to desired policies), and the ability to effectively process information in favor of these policies (and other longer-term possibilities). 

The variables modulating regimes of attention that have practical modeling potential are plenty, and in sections 4 and 5, our interest is in the control, influence, and synchronization of attention, which lead to social power structures. Our focus will be on empowerment quantified as agents’ potential influence over their environment (which is tracked by assessing the mutual information between actions and states). We will analyze this through formal methods that are tractable, but this presents us with indications of future modeling possibilities, which in our context remain unformalized as they are practically intractable. We can think, for example, about the modeling of “inheritance” or “provenance” variables: wherever/however an agent exists, they could be born into a rich family, in a specific part of the world, etc., and these factors have an influence on how others interpret them (and therefore on life-expectancy: mortality, and many other crucial effects). One can also posit a possible “frictions/contingencies/contextualities” variable—in colloquial terms “you have to work with what you've got”—by which, even if agents can attract others’ attention and process information effectively towards the effectuation of a specific policy, the projected goal and thus possible state space is constrained by a context which can be highly divergent from the “ideal” one they project as required for bringing a goal about. Confidence in policies is a well-framed and studied variable in the context of AIF, and in the literature it is often directly linked to the concept of power: “self-confidence [is] the person’s experience of their power to act in the world” \citep{kiverstein2019obsessive}. There are more, of course, and it depends on how granular one may want to go, these are examples which give the reader an indication of our intentions in later sections.

Power as potential, as force: as the ability to do work in physics and the ability to do work in the socioreproductive realm, represent two sides of the same coin: the mechanics of matter and energy sustain the dynamics of complex systems; which, in turn, modulate matter and energy. Under AIF, this ‘common currency’ can be studied as the process by which state spaces are explored by collectives (dis)attending to projective possibilities. In order to understand this better, formalizations are necessary, which is our interest in this paper: awareness about the modulation of attention begins at the level of understanding the constraints framing attention. As Foucault notes in “The Subject and Power” \citep{foucault1982subject}, formalizations are not without their reductive, objectifying problems, but are necessary in order to drive critical analysis forward. Further, he also notes: “What we need is a new economy of power relations—the word “economy” being used in its theoretical and practical sense” (p. 779). This is, in great part, what we aim to explore here, not an established structure but “a technique, a form of power” (p. 781): attention itself, analyzed under AIF. 

In the socioconstructive realm we currently observe: those under duress resort to inevitably short-term policies, as well as pool their attentional trust in those who seem to promise them less uncertainty\footnote{“...in trials in which the agents’ prior beliefs are strong, such as after a change of mind, they do not simply follow one another, but try to fulfill their prior belief—and this explains why we observe several errors after only one of the agents changes mind. These examples illustrate that it is the strength (or the precision) of the beliefs about the joint goal context that determines whether or not an imitative response takes place.” \citep{maisto2024interactive}.}. Those who can rely on the minimization of uncertainty because of historically-given, infrastructural factors have access to more and therefore longer-term leverage on the unfolding future. Power as the exertion of longer-term agency implies that, to begin with, an agent relies on factors that make more spacetime available for the computation (as projection: planning and effectuation) of possible futures.

We now present the theoretical framework before moving on to defining formal aspects of this, i.e., the ways in which we may organize variables in narratives as scripts, and how controlling a narrative or exerting a script implies controlling attention, which implies power.

\section{Theoretical Framework: Active Inference}

AIF posits that biological systems maintain their existence by minimizing a quantity known as free energy—an upper bound on surprise or unlikeliness given sensory data. AIF extends this principle to explain how agents—whether cells, brains, or organisms—actively align their internal models with their environments to ensure adaptive behavior. It ascribes a dual role to perception and action: while perception updates beliefs about the world based on sensory inputs, action minimizes discrepancies between sensory inputs and predictions by altering the environment to match internal expectations (Friston, 2009).

This is fundamentally embodied and enactive. Through it we integrate ideas from predictive processing and Bayesian inference. It is a useful tool to cast self-organizing systems as nested within hierarchical generative models that predict sensory inputs at multiple spatiotemporal scales. 

Interestingly, we can also link it to an enactive inference perspective, which focuses more strongly on the role of action in shaping perception. Under this lens, beliefs are formed actively. For readers unfamiliar with the intricacies of Active Inference Framework (AIF), this concept can be seen as a dynamic process where agents optimize their interactions with and models of the world to reduce uncertainty and maintain homeostasis, allostasis, or homeorhesis \citep{ororbia2023mortal}. AIF highlights the importance of generative models that represent the structure of the environment and guide behavior. It directly integrates perceptual, cognitive, and motor processes.

Generative models are thus the heart of any AIF agent. They represent a structured way for agents to analyze sensory inputs and infer their possible causes. These models consist of hierarchical representations that generate predictions about sensory states based on beliefs (i.e., inductive habits) about hidden causes. Generative processes represent the true states of the world that produce sensory observations, while generative models represent the agent's internal hypotheses of these processes. AIF hinges on the alignment of models with processes, with actions taken to correct any mismatch. This framework allows agents to learn both the structure of the environment and effective policies through message-passing algorithms, which mimic neuronal processes \citep{parr2019neuronal}.

Generative models can be organized hierarchically, with each layer producing predictions about the level below it. For example, higher levels might model abstract causes (such as intentions or environmental conditions), while lower levels represent more immediate sensory or motor states. Each layer generates predictions (i.e., expected sensory states) based on inferred hidden causes and corrects these predictions based on incoming data. This process is mathematically described through conditional dependencies that cascade through each layer.

In generative models, the primary components involved in sense-making include observable sensory states, hidden states, and parameters that encode the probabilistic dependencies between them:
Sensory states are the inputs that the agent receives from the environment. They are the observations the model is trying to predict or explain based on hidden states and other parameters. Sensory states provide feedback to the generative model, helping the agent assess how well its predictions align with real-world data. This can be represented as $P(o \mid s, \theta)$, where $s$ is a hidden state, and $\theta$ represents the parameters of the generative model.
Hidden states capture the unobserved causes or latent variables that the agent infers to explain sensory states. Hidden states can represent underlying factors or situations that produce the sensory input, which aren’t directly measurable but can be inferred from patterns in sensory states. In Bayesian terms, these are inferred by the model to reduce uncertainty (or surprise) about observable data. The distribution over hidden states can be represented as $P(s \mid \theta)$, where $\theta$ (i.e., the generative model parametrization) influences the hidden state predictions based on prior beliefs or learned parameters.
Parameters define the relationships between hidden and sensory states. Parameters capture the structural and statistical dependencies within the model and are updated over time to maximize model fitness to the data. Parameters specify the exact form of dependencies, such as the strength or variance of the connection between hidden states and observable outcomes.

The likelihood function defines the probability of observations $o$ given hidden states $s$ and parameters $\theta$. It measures how well a specific combination of hidden states and parameters can predict or explain the observed data. The \textbf{prior distribution} encapsulates the model’s beliefs about the hidden states before observing any data. This distribution is essential in Bayesian inference as it provides a reference point or initial expectation about the hidden states. The \textbf{posterior distribution} represents the updated belief about hidden states after observing sensory data. Using Bayes’ rule, the posterior distribution combines the likelihood and prior to form an updated probability distribution over hidden states.

Model evidence, or marginal likelihood, represents the probability of observing data under the generative model, integrating out the hidden states. This term is used in model comparison and is a measure of how well the model as a whole can explain observed data.

\subsection{Free energy minimization}

Free energy minimization is the unifying principle that motivates perception, cognition, and action in Active Inference Framework (AIF). As described in Parr and Friston's 2019 paper, this minimization occurs at both local and larger scales \citep{parr2019neuronal}. Free energy minimization at cellular and cognitive levels could reflect larger systemic constraints and efficiencies. This process involves reducing the \textit{divergence} between an agent's predictions and its sensory inputs, maintaining a state of low surprise or entropy.

Free energy is an upper bound on the ``surprise'' (or negative log-probability) an agent experiences given its observations ($o$). Surprise reflects the degree to which sensory data diverges from an agent's predictions, but it is generally intractable to compute directly. \textit{Variational free energy} (VFE), an approximation applied by Friston and others to the study of the brain \citep{friston2006free,friston2015active,friston2016active}, allows the agent to sidestep this intractability by using an internal generative model, represented as $q$-distributions over hidden (latent) states ($s$), to approximate the posterior distribution. The variational free energy $F$ can be expressed as:

\begin{align*}
    F &= \underbrace{\DKL[q(s) \, \| \, p(s \mid o)]}_{\text{Divergence}} - \underbrace{\ln p(o)}_{\text{surprise}}\\
    &= \underbrace{\DKL[q(s) \, \| \, p(s)]}_{\text{Complexity}} - \underbrace{\mathbb{E}_{q(s)}[\ln p(o \mid s)]}_{\text{Accuracy}}.
\end{align*}

Here, $\DKL[q(s) \, \| \, p(s \mid o)]$ is the Kullback-Leibler (KL) divergence between the approximate posterior $q(s)$ and the true posterior $p(s \mid o)$. Minimizing this term aligns the agent's beliefs (encoded in $q$) with the posterior distribution. 

The actual sensory data (encoded in $p$). The $-\ln p(o)$ term represents the surprise, or the negative log model evidence, which represents how poorly a model fits the data it is trying to explain.

$\mathbb{E}_{q(s)}[\ln p(o \mid s)]$ represents the expected log-likelihood of observations given the hidden states. Maximizing this term makes the agent's model predictive of actual sensory inputs. To minimize free energy and align their sensory inputs with their predictions, agents have to update their beliefs about the world. Belief updating under AIF is achieved through variational Bayesian inference, whereby agents continuously update their beliefs about the state of the world based on new sensory information. This allows agents to make predictions and adjust their actions to confirm these predictions, thereby preserving a coherent sense of self and environment. 

Given appropriate belief updating, the system can then select actions that are consistent with its internal generative model. This effectively ensures that the chosen policies minimize future free energy. This mechanism has broad implications, enabling goal-directed behavior and coherent action selection, even in novel environments.

Minimization can be broken down into two complementary strategies. Agents can perform \textit{perceptual} or \textit{active} inference, which are inextricably related. Perceptual inference entails adjusting beliefs about hidden states to make them more consistent with current sensory data, which is akin to learning or updating internal representations to match the environment. AIF, on the other hand, entails \textit{modifying} the environment or seeking new sensory data to make incoming data consistent with beliefs, guiding action selection and resulting in a feedback loop where actions bring sensory data closer to expectations.  It is important to note difficulty in differentiation between the active and the perceptive, which brings us to the question of modeling collective desire: is there a way to tell the difference between when we act according to an “inner” logic we have conscious cognitive access to, and when are we moved by a logic we simply ‘feel’ driven by? Friston and others postulate that free energy minimization, as a principle, might reflect broader systemic efficiencies in biological scaling laws, where living systems seek to maintain optimal internal states (allostatic setpoints) with minimal energetic cost \citep{vasil2020world}. The “average” aspect is important to emphasize here: for complex systems, long-term averages may signify significant shorter-term high-risk, high-energy investments.

For example, cells engage in biochemical processes that aim to maintain homeostasis, and deviations from expected internal states (e.g., due to external changes) trigger corrective actions, such as upregulating certain pathways to maintain ion balances or in response to increased energy demand. This is a form of local free energy minimization. In cognition, the brain applies predictive coding, where sensory inputs are compared against predictions generated by a hierarchical generative model. Mismatches trigger updates (prediction errors) at successive hierarchical levels, ultimately influencing cognition and behavior to minimize surprise. Free energy minimization translates into entropy reduction. Entropy here quantifies uncertainty about the states of the world, and free energy minimization reduces this uncertainty by aligning beliefs (internal model) with reality (sensory inputs).

Part of the internal model of an agent may include predictions about its own actions. Such predictions can be thought of as \textit{plans}, which may be optimized to bring about desirable outcomes. In information-theoretic terms, this results in a shift from minimizing the Variational Free Energy (VFE) to minimizing the \textit{Expected Free Energy} (EFE), which can be decomposed into two components—\textit{extrinsic value} (goal-directed, rewarding outcomes) and \textit{epistemic value} (uncertainty reduction). Minimizing EFE addresses both immediate needs and long-term information gathering, balancing exploration and exploitation through actions aimed at reducing uncertainty about future outcomes. It is expressed as:

\begin{align*}
    G = -\underbrace{\mathbb{E}_{q(s,o \mid \pi)}\Big[ \DKL[q(s \mid o, \pi) \| q(s \mid \pi)] \Big]}_{\text{Epistemic value}} - \underbrace{\mathbb{E}_{q(o \mid \pi)}[\ln \tilde p(o)]}_{\text{Pragmatic value}},
    % G = \mathbb{E}[\DKL(q(s') \| p(s' \mid a))] + \mathbb{E}[U(s')]
\end{align*}

where $\ln \tilde{p}(o)$ represents the agent’s preferences over observations or \textit{extrinsic value}, expressed as a (biased) prior probability distribution over observations \citep{parr2022active}. The first term represents the (negative) \textit{expected information gain} or \textit{epistemic value}, which scores the degree to which an agent expects to update their beliefs in light of new sensory data. By pursuing outcomes that maximize a balance between epistemic and extrinsic values, agents simultaneously satisfy immediate goals and resolve uncertainty, driving adaptive behavior in complex environments.

\subsection{Why active inference is suitable for modeling power dynamics}
Active Inference Framework (AIF) is well-suited to understand power dynamics, beginning with individual agents as predictive systems that minimize uncertainty through perception and action \citep{bruineberg2017active}. These agents develop control through embodied interaction with their environment, establishing what Bruineberg calls the ``primacy of the 'I Can'''—the fundamental ability, or \textit{power}, to influence one’s surroundings. This individual control scales up through generative models that enable agents to predict and understand others’ intentions \citep{friston2020generative,vasil2020world}, while attention is fundamentally shaped by deep preferences that determine what becomes salient in the environment \citep{parvizi2024preferences}.

At the social level, power dynamics emerge through shared predictive models and environmental mediation. Groups develop shared narratives that coordinate actions and expectations \citep{bouizegarene2020narrative}, while stigmergic processes allow for distributed control through environmental modifications that influence collective behavior \citep{friedman2021active}. This social coordination is reinforced through \textit{deontic} cues that signal social norms and expectations \citep{constant2019regimes}, while cultural practices shape attention patterns to create shared perceptual landscapes within the environment. Joint attention and shared fields of affordances enable efficient coordination through mutual understanding of action possibilities \citep{tison2021communication,vasil2020world}.

These mechanisms ultimately can give rise to complex power structures through decentralized interactions rather than top-down control \citep{friedman2021active}. Groups self-organize into subgroups based on belief similarity \citep{kastel2023small}, while influence spreads through network structures that amplify certain messages while marginalizing others \citep{albarracin2022epistemic}. This process creates systems of ‘cultural capital’ \citep{bourdieu2018structures} where shared knowledge and norms enable efficient navigation of social contexts \citep{constant2019regimes,hipolito2022enactive}. The result is a dynamic system where power emerges from the interplay of individual prediction, social influence, and cultural learning, all of which can be unified under the principles of AIF.

The question of social sameness (as in “same” narratives, belief similarity, etc.) and (dis)attention to it, is therefore a political question. If attention is fundamentally shaped by deep preferences that determine what becomes salient in the environment for an agent or group, then we ought to ask how the production and performance of scripts (habitus, narratives: histories, identities, etc.) can be newly understood so that we may critically construct these collectively, without risking that this collectivity is induced by strong homogenizing scripts which occlude or incapacitate ever-renewing possibilities (as warned by Arendt). Incentives to promote legibility in a population can also manifest in institutions shaping and direction of attention toward specific narratives \citep{scott2020seeing}, and may similarly arise in collectively determined scripts. Thus, a balance needs to be maintained between benefits such as improved coordination and an oversuppression of diversity among constituent members of collectives. As mentioned earlier, “There is no alternative” can be understood as the poster-child of ‘one-dimensional’\footnote{For this Marcusian notion at work in similar ways in the domain of NLP, see \citep{dejager2023semantic}.} visions effectuated by reducing complex systems into simplified structures which pool attention towards a specific goal.  Through our treatment, a new definition of social power can thus be understood as possibilistic. Essentially: if we ought to adopt a politics based on the knowledge that scripts work in the ways discussed above and in what follows below, and if the organization of life is hierarchical by definition (both physically and metaphysically: we are given to a world of already-existing constraints which we organize around by means of concepts aimed at future possibilities), taking account of this by understanding the nature of distributed generative models—over time: as generations, as the representation of historical narratives, etc.—offers a vision of possibilistic landscapes based on collective dynamics (and their analyzable mechanics) which can be framed beyond our current visions based on simplistic ideas of agency as individualized freedom and through atomized actors. 

\section{Social Scripts and Power}
\subsection{Scripts as frameworks for social behavior}

Social scripts are blueprints that enable agents to navigate social situations and select situationally appropriate behaviors \citep{albarracin2021variational}. They represent both internalized cognitive schemas and externalized social orders that structure behavioral possibilities within cultural contexts. Scripts harness socially shared knowledge that prescribes normative standards for behavior, similar to how actors share a dramaturgical script.
Strong scripts specify precise sequences of structured behavioral events that agents are expected to perform in certain social situations \citep{albarracin2021variational}. These scripts maintain causal and temporal relationships between events, enabling agents to make reliable inferences about others’ likely actions. For example, the North American restaurant script prescribes specific ordered sequences: being seated by the host, reviewing menus, ordering drinks, ordering food, eating, and paying the bill including tip \citep{albarracin2021variational}.

Weak scripts specify the typical features and semantic associations that agents encounter in particular event types, without strictly prescribing event order \citep{albarracin2021variational}. These scripts function as clusters of conceptually related possibilities that help agents interpret situations and identify appropriate actions. The semantic connections in weak scripts adapt agents’ perceptual fields by making certain action possibilities more salient based on contextual goals.

Scripts help maintain power structures. Scripts normalize certain behavioral patterns by providing templates for "appropriate" action, making deviations appear abnormal or inappropriate (Constant et al., 2018). This normalization reinforces existing power relations by making them appear natural and inevitable. They establish reliable sequences of interaction in favor of a current norm: whether agents following this are (critical) aware of the norm, or simply perform it, is often the question. They thus reduce uncertainty and create predictable social patterns that benefit those already in power \citep{albarracin2021variational}. These patterns become self-reinforcing as agents learn to expect and reproduce them. Given this predictability, scripts assign and maintain social roles by specifying appropriate behaviors for different agents based on their position within power structures \citep{} (Constant et al., 2018). This role differentiation helps reproduce hierarchical relationships.

Looking at scripts under AIF, they can be understood as shared generative models that enable coordinated social behavior. Scripts represent shared knowledge about the causal structure of social situations, allowing agents to make predictions about others' behavior and select appropriate actions. These shared models facilitate social coordination by reducing uncertainty about interaction patterns. Scripts incorporate deontic cues that trigger specific behavioral policies, helping agents select appropriate actions in response to environmental signals \citep{constant2019regimes}. These cues guide action selection by making certain policies more likely in specific contexts. Scripts are transmitted across generations through cultural learning processes, enabling the reproduction of social structures. This transmission occurs through both explicit instruction and implicit learning from environmental regularities.

\subsection{ Narratives as Scripts}
The role of narratives in power dynamics, as scripts which normalize behavioral patterns by providing templates for ``appropriate'' action, may make ``deviant'' behavior appear explorative (others may learn, inspired by their curiosity) or inappropriate \citep{constant2019regimes}: leading to social effects such as castigation or neglect. This leads to the creation of shared meaning as well as to the justification of existing power structures. Because of the challenges inherent to the coordination of social reality—as mentioned: whether agents following a script are (critically) aware of the implied norm, or simply perform it, being the main question—we can also think about scripting narratives in terms of self-fulfilling prophecies, hype, fictions that make themselves real, etc.: none of the possibilistic options on social offer need be realistic, so long as they are strong enough to attract attention (driven by the desire to reduce uncertainty about future states).

The effects of variational free energy-minimization are observed at the level of the basic structures of language: (semantic) attractors such as redundancy, rhyme, alliteration, etc., can be understood as patterns which aid memory, communication and information retrieval policies, because they permanently reappear, pooling attention to the “same” phenomena, thus greatly reducing the need to interpret and organize linguistic interaction and cognition anew. We can understand, for example, the possible emergence of rhythmic patterning in communication as enabling the storage and access of information collectively: by connecting a sound-pattern to a phenomenon, we reduce uncertainty about reality once that sound-pattern is witnessed, in true Pavlovian fashion. By way of the same uncertainty-minimizing mechanisms, (simple or complex) narrative constructs create cognitive situations where agents reduce complex aspects of the world down to preferred scenarios: it is easier to imagine the end of the world than the end of capitalism \citep{fisher2009capitalist}. In light of our arguments: this might be, perhaps, because the end of the world is relatively easy to imagine: asteroid collision, megavolcano, etc., while the systemic nature of capitalism is rather difficult to pull apart in detail. Or: it is easier to reduce uncertainty about my current stress-predicaments by blaming my neighbor’s lifestyle, than by reassessing my entire lifestyle (generative model) and change it on the basis of a myriad other variables. Cognition, often led by a desire for and mechanics of the shortest path, tends towards predictable, digestible narratives more often than not.

Unlike short-range projective experiences such as the ones detailed in \citep{maisto2024interactive}, where prediction errors present immediate challenges, narratives and scripts provide vastly speculative semantic spaces for exploring models against hypothetical outcomes. As we have been highlighting, with regard to power, the problems emerge in situations where agents controlling dominant narratives are guided by nothing other than power-acquisition or maintenance, leading not only to, e.g., misinformation, echo chambers, conspiracy theories, etc. \citep{albarracin2022epistemic}, but, in political communication and organization, to the justification of future policies based on (supposedly) given conditions (deriving an ought from an is, essentially). 
Abstract semantic constructs (fictions, narratives, scripts), which can be understood as constraint-regimes coming to operate as distributed generative models, create cognitive situations where agents can learn by “internally” simulating responses to high-stakes scenarios, while maintaining corporeal integrity and safety, all the while plastically exploring their epistemic condition. An environment pooling attention to strong-enough narratives will elicit the scripting of policies in line with those narratives (as mentioned, regardless of their ‘veridical’ content). For agents under thermodynamic duress (socioeconomic, corporeal, etc.): the simpler the narrative, often, the better. This being for the banal reason that they are under duress, and therefore incapable of making enough spacetime available for deep, projective computations. Narrative structures tempered by the pooling of attention act as semantic attractors which organize possible world-states into explorable patterns. These patterns function as probability-possibility landscapes for belief-updating, allowing agents to or restricting them from fine-tuning their predictive models through engagement. Considering homo sapiens’ universal tendency towards narrative-sharing, this process can be understood as a corollary of actively-inferential social agents, particularly if we understand ‘control’ as the phenomenon of conscious attention to how the modulation of variables effectuate the production of the future\footnote{ In the context of “fictional inference” \citep{dejager2023semantic}, which is much of what humans do, this is the exploration of domains where direct experience would be prohibitively risky or even impossible. Fiction provides “all of the learning, at relatively low risk” (ibid.).}.
The theoretical framework developed here has significant implications for understanding both the cognitive foundations of narrative engagement and how the power of attention to these results in the scripting of communal behavior. We now move to an analysis of power through the lens of AIF.

\section{Power Through the Lens of Active Inference}
\subsection{Individual agency and empowerment}

Autonomy refers to a system's capacity to govern itself based on internally defined norms \citep{watson2005autonomous}. Autonomous behavior emerges through entropy maximization under constraints. Agents do not simply maximize randomness (though in some instances they do: explorative behavior is subject to high noise and benefits from high noise, again, note on average long-term VFEM in social agents). Agents optimize their behavior by maintaining a balance between maximizing entropy and respecting system constraints \citep{jaynes1957information}. This is consistent with Jaynes' principle of maximum entropy, which states that the least biased estimate of a probability distribution is the one that maximizes entropy subject to known constraints. In autonomous agents, these constraints include both the agent's internal model and environmental dynamics \citep{golan2022understanding}.

We can think of \textit{empowerment} as a formal measure of an agent's control over its future states, defined as the maximum mutual information between an agent's actions and the resulting states \citep{klyubin2005empowerment}:
% \[
% E = \max_{p(a)} I(A; S)
% \]
% or equivalently:
\[
E_{t,n} = \max_{p(a_{t,n})} I(A_{t,n}; s_{t+n}),
\]
where $I(A_{t,n}; s_{t+n})$ denotes the mutual information between the next $n$ actions $A_{t,n}$ from time $t$ and the state $s_{t+n}$ at time $t + n$, and $p(a_{t,n})$ represents the probability distribution over the next $n$ actions from time $t$ \citep{klyubin2005empowerment}. In this context, agents select actions that maximize their control over future environmental states, optimizing their ability to influence outcomes. Higher empowerment indicates greater potential for an agent to achieve desired future states. Importantly, empowerment can be decomposed into entropy terms as:
% \[
% I(A; S) = H(A) + H(S) - H(A, S)
% \]
\[
I(A_{t,n}; s_{t+n}) = H(A_{t,n}) + H(s_{t+n}) - H(A_{t,n}, s_{t+n}),
\]
where $H[X]$ represents the Shannon entropy of the random variable $X$ and $H[X, Y]$ denotes the joint entropy of $X$ and $Y$. This decomposition reveals that empowerment involves maximizing the marginal entropies of state and action distributions while maintaining strong correlations between actions and their consequences \footnote{We would like to thank Alex Kiefer for pointing out this breakdown of empowerment in terms of entropy, and for suggesting the broader idea of casting intrinsic motivation as constrained entropy maximization.}. We can also link these ideas to recent developments in understanding intrinsic motivation in dynamical control systems \citep{tiomkin2024intrinsic} and complex behavior emerging from the drive to occupy future action-state path spaces \citep{ramirez2024complex}.

Empowerment as defined above is an \textit{intrinsic} motivation that does not specify how an agent should take actions to achieve specific goals. Extensions have been introduced to incorporate goal-directedness by introducing a reward constraint \citep{volpi2020goal} or tradeoff \citep{leibfried2019unified}, which highlight the tensions that may arise between intrinsic and extrinsic motivators.

Under the Active Inference Framework (AIF), autonomous agents select policies (sequences of actions) by minimizing expected free energy, which includes both epistemic (information-seeking) and pragmatic (goal-directed) terms \citep{friston2010action,parr2022active}. AIF differs from empowerment in several respects, but one crucial point of difference is that it explicitly models the partial observability inherent in many decision-making environments, motivating agents to resolve uncertainty through curiosity-driven behavior. This balance between exploration and exploitation allows agents to both gather information about their environment and pursue specific goals. Policy selection in autonomous agents involves maintaining a generative model of environmental dynamics. To do so, agents must update their beliefs about current states through perceptual inference. Then, they can select actions that minimize expected free energy over future states and ultimately learn from action outcomes to improve the generative model.

Both empowerment and the EFE highlight different components of power that are relevant to the present discussion. Empowerment more directly formalises the notion of an agent's control over its environment, while the EFE emphasizes the role of information gain in the context of generative models, pointing to the importance of understanding how informational landscapes and affordances can be shaped by those who wield social power. Developing a unified framework for integrating these different objectives would likely provide a deeper understanding of how power, goals, and information relate to one another.
% Agents can simultaneously maintain control over their environment while remaining adaptable to changing circumstances.

\subsection{Social power dynamics}
Social power dynamics emerge as a function of interacting perception-action loops across multiple agents. Power is the ability to shape both the physical and informational environment. Power operates as an attractor in social systems, creating a force that draws agents toward centers of influence. This gravitational effect works through two primary mechanisms. Entities with high power gravity naturally draw others to them through access to resources, information, or strategic advantages. As more agents are drawn into their ‘orbit,’ their influence amplifies through positive feedback loops, e.g., through phenomena such as preferential attachment.

Powerful entities offer perceived stability and predictability in an uncertain environment. In AIF terms, they help other agents minimize their expected free energy by providing reliable frameworks for prediction, essentially by dictating the information geometry landscape \citep{constant2019regimes}.
Power confers significant advantages in information processing and management. Powerful entities occupy positions that grant them privileged access to information streams, enabling more accurate modeling of their environment. Ultimately, they also control what can and cannot be modeled, and in what way, through their position and resources, powerful agents can offload computational demands onto others or their environment, allowing for more efficient processing of complex information. In the complexifications of the “free market” condition unfolding over the 20th century, we can understand these dynamics as the emergence of a novelty-generating network of agents which has proven to provide benefits to those with leverage over its structures: agents already well-positioned within the informatic domain. In sociological lingo: capitalist capture, or the cancellation of the future \citep{fisher2009capitalist}. Power influences how errors and uncertainty are processed within social systems. Powerful entities can better absorb or reinterpret errors due to their broader resource base and influence over narrative construction. Through their ability to shape both physical and informational environments, powerful agents can actively reduce uncertainty for themselves while potentially increasing it for others.

The manipulation of information geometry—both in terms of perception and action—is central to how power operates in social systems. Powerful entities can shape the attentional landscapes of others, directing focus and resources toward certain states while making others less salient or accessible. This ability to manipulate both the physical and informational structure of the environment creates a self-reinforcing cycle where power begets more power through increasingly sophisticated control over the collective predictive landscape. Power also expands an entity's available state-space policies in several ways: powerful entities have access to a broader range of possible actions and strategies through their resources and influence. Through their attractive force, they gain access to the state-space policies of agents within their sphere of influence, effectively expanding their operational capabilities.

\subsection{Formal Framework Integration}

Power dynamics within and between agents can be formalized by addressing various dimensions of control, influence, and synchronization. At the individual level, recall that \textit{empowerment} quantifies an agent’s potential influence over its environment by assessing the mutual information between the agent’s actions and the resulting states \citep{klyubin2005empowerment}. Here, we discuss some implications of this notion in the context of active inference.
% This is captured mathematically as:

% \[
% E_{t,n} = \max_{p(a_{t,n})} I(A_{t,n}; S_{t+n}),
% \]

% where $I(A_{t,n}; S_{t+n})$ denotes the mutual information between the next $n$ actions $A_{t,n}$ from time $t$ and the state $S_{t+n}$ at time $t + n$, and $p(a_{t,n})$ represents the probability distribution over the next $n$ actions from time $t$ (Hynes, 2024). In this context, agents select actions that maximize their control over future environmental states, optimizing their ability to influence outcomes.

% The underlying principle guiding these dynamics is the minimization of \textit{variational free energy} $F$, the principle governing the cyclic unfolding of perception and action. Variational free energy is expressed as:

% \[
% F = \mathbb{E}_{q(s)}[\ln q(s)] - \ln p(o, s),
% \]

% where $\mathbb{E}_{q(s)}[\ln q(s)]$ represents the entropy of the approximate posterior, and $\ln p(o, s)$ represents the joint probability of observations and hidden states.
% where $Q(s)$ is an approximate posterior over states and $P(o, s)$ is the generative model, linking observed outcomes to their latent causes (Constant et al., 2018). The minimization of  free energy maximizes the evidence for agents’ internal model of the world, guiding perception and action toward alignment with their environment.
In social contexts, free energy is expanded to incorporate \textit{deontic value}, which reflects the influence of social norms and expectations on behavior. This is formalized as:
\[
P(\pi \mid o) \propto \exp(-G(\pi) + D(\pi, o)),
\]

where $G(\pi)$ represents expected free energy for a policy $\pi$, and $D(\pi, o)$ is the deontic value associated with aligning actions with social norms and expectations \citep{constant2019regimes}. Social norms modulate policy selection, encouraging behavior that is goal-oriented and aligned with collective expectations. This deontic value may also be linked to a particular kind of empowerment of other agents. Consider that to another agent, one's own actions may be viewed as part of the external state that is being predicted and controlled. Therefore, the empowerment of other agents includes their influence on one's own policy, which may be reflected by the deontic value term above. A deeper investigation into how the two are precisely related is beyond the scope of this paper, but would shed light on the relationship between social norms and individual/collective empowerment.

Beyond control over other agents' policies, agents may also exert influence over others' beliefs, constituting another aspect of empowerment which focuses on viewing other agents' generative models as states external to oneself. In particular, one phenomenon worth investigating is the synchronization of beliefs between agents in social interactions. This can be quantitatively assessed through the \textit{Kullback-Leibler divergence} between different agents' generative models. In the simple case of two agents $i$ and $j$, this can be quantified as:
\[
\DKL[Q_i(s) \| Q_j(s)] = \mathbb{E}_{Q_i}[\ln Q_i(s) - \ln Q_j(s)],
\]
where $Q_i(s)$ and $Q_j(s)$ denote the belief distributions of agents $i$ and $j$ over the hidden state, respectively. This quantity reflects the degree of belief alignment, indicating the extent to which a \textit{shared} understanding emerges from social interaction. The ways in which this is unconscious, tacit, or otherwise unexplored are the elements which lead to many of the intractable aspects of formalizing these dynamics. This is a consideration which ought to be explicitly stated in any modeling attempt.

% Belief alignment itself can emerge from reciprocal interactions governed by the minimization of \textit{expected free energy} $G$, which balances uncertainty reduction with goal-directed actions. For a given policy $\pi$, expected free energy can be decomposed as:

% \[
% G(\pi) = \mathbb{E}_{Q(o, s \mid \pi)}[\ln P(o, s \mid \pi) - \ln Q(s \mid o, \pi)].
% \]

% where $P(o \mid s, \pi)$ is the likelihood of outcomes given states under policy $\pi$, and $Q(s \mid o, \pi)$ is the posterior beliefs about states based on outcomes. The first term, an intrinsic uncertainty-reducing component, encourages agents to seek out informative outcomes, while the second term, representing extrinsic value, aligns behavior with desirable goals (Friston et al., 2020).

Belief alignment itself can emerge from reciprocal interactions governed by the minimization of variational free energy. 
% In social exchanges, as agents update their beliefs in response to outcomes generated by each other, it may be the case that parts of their generative models become synchronized to each other, thus establishing a shared narrative or even sense of identity. 
This may arise as agents select actions consistent with their internal models, and thus produce outcomes that align with their beliefs, which reciprocally influences the beliefs of their counterparts. Over successive interactions, the minimization of expected free energy can lead to synchronized beliefs (encapsulated by the KL divergence), by a process of mutually adapting beliefs and behaviors in order to achieve benefits from coordination, which includes both pragmatic and epistemic components\footnote{ In adversarial or mixed-motive settings, this synchronization need not occur, e.g., in situations where an agent is being deceived or manipulated by another.}. Repeated exchanges effectively reduce this divergence:
% \[
% \lim_{t \to \infty} \DKL[Q_i(s) \| Q_j(s)] \approx 0.
% \]
\[
\lim_{t \to \infty} \DKL[Q_i^t(s) \| Q_j^t(s)] \approx 0,
\]
where the superscript indicates the generative model of each agent at different points in time. Intuitively, under appropriate interaction conditions, agents share beliefs (typically in the form of likelihoods) with one another and incorporate these into their own generative models, thus bringing their states of belief into alignment with each other. This signifies a stable convergence where the beliefs of both agents synchronize, establishing a shared understanding and a common ground for interaction.
There are many ways this synchronization can occur, with extreme cases being where 
\[
\lim_{t \to \infty} \DKL[Q_i^t \| Q_i^0] \ll \lim_{t \to \infty} \DKL[Q_j^t \| Q_j^0],
\]
or vice versa. Intuitively, these correspond to situations where one agent makes a small update to their beliefs but causes the other to change their beliefs to a greater extent to fit their own. The agent who makes the significantly smaller update may do so as a result of being in a position of greater social power.

Linking this idea back to \textit{empowerment}, notice that defining ``external states'' is more complicated in social contexts, because one agent’s internal state is another agent’s external state, and vice versa. This means that different agents’ internal representations of the world must be similar in some parts (those that are external to all of them) and different in other parts. Therefore, part of what it means for an agent to maximize their empowerment in a social context is to actively make parts of other agents more predictable and useful to themselves, which is often more easily achieved by bringing other agents’ generative models in line with one’s own.
In general, belief alignment can emerge as a natural consequence of each agent’s drive to minimize free energy within the right social context. This fosters synchronized behavior, shared narratives, and collective understanding, unifying individual agency, social influence, and collective behavior under the cohesive formalism of AIF \citep{friston2020generative,friston2024designing}. However, as the brief analysis above reveals, such synchronization through belief alignment may come about as a result of the exertion of social power, highlighting the importance of understanding not just when this synchronization occurs, but also how it does so.
As we conceptualized it, power is an agent's capacity to shape and control the (semantic) ``attractors'' within a social or informational landscape. These attractors represent stable states or patterns that agents are drawn towards, which reduce uncertainty and create predictable social structures. Power enables agents to influence which states become salient or preferred for others, thereby controlling the flow of attention, shaping beliefs, and modulating the precision of predictions. The 20th and 21st century have shown, time and time again—especially so in the last 20 years—how mediatic domination (narrative control) results in the shaping of attractors which organize life.

An agent with power has the capacity to shape these expected outcomes by adjusting the social or informational environment in a way that amplifies certain states while suppressing others. For example, powerful agents may create narratives, i.e., scripts, that reduce the dimensionality of the state space by establishing strong, predictable attractors—such as social norms or shared narratives. By doing so, they effectively direct attention and ensure that other agents align their actions with these preferred states.

\[
\pi^* = \arg\min_{\pi} G(\pi) + \alpha \cdot \text{precision}(Q(s \mid o, \pi)),
\]

where $\alpha$ is a modulating factor that reflects the power to affect the perceived reliability of beliefs, amplifying the impact of certain beliefs and actions across a social network.

% The ability to direct attention through attractors and precision modulation defines power in AIF. 
% This alignment process can be mathematically formalized using the \textit{Kullback-Leibler (KL) divergence}, which measures the difference between the belief distributions of two agents. For two agents $i$ and $j$, the KL divergence is defined as:
% \[
% \DKL[Q_i(s) \| Q_j(s)] = \mathbb{E}_{Q_i}[\ln Q_i(s) - \ln Q_j(s)],
% \]
% where $Q_i(s)$ and $Q_j(s)$ represent the beliefs of agents $i$ and $j$, respectively, about a particular state $s$.
% A lower KL divergence value indicates a closer alignment of beliefs.
In leader-follower dynamics, a leader minimizes this divergence by exerting control over precision, amplifying others' confidence in specific interpretations of their sensory data. For example, if the leader’s goal is clearly defined and they select social epistemic actions (more informative, albeit less efficient actions), they convey their intentions early and reduce uncertainty for followers \citep{maisto2024interactive}. This approach encourages other agents to update their beliefs to align with the leader’s, effectively lowering KL divergence across the group.
Even in cases where leadership is not predefined, an agent can assume a leadership role by adjusting precision to show high confidence in their beliefs or goals. This increase in precision draws other agents toward the confident agent's stance, resulting in a convergence on shared beliefs. Precision-driven leadership can thus promote belief alignment without coercion, as agents gradually follow those who express strong, reliable predictions. In this way, the power to direct attention, structure narratives, and therefore modulate precision enables agents to actively shape collective beliefs, creating cohesion within groups and guiding group behavior through the minimization of belief divergence.

The applications of this inevitably begin at reductive formalizations (and the math is never the territory \citep{andrews2021math}), however, the implications of these arguments for complex social dynamics in the realms of, e.g., policy and governance are vast. Coordination dynamics at the level of simple perception-action cycles such as the ones analyzed in \citep{maisto2024interactive}, need to be framed in terms of their possibilities for novel understandings of social change processes: how should humanity—as one, as many—develop strategies with enough coherence and plasticity so as to resist the oppressive aspects of dominant narratives which do not benefit the coordination of the whole? What kinds of timeframes are we thinking about (in terms of generations, or the reorganization of our understanding of future time altogether)? Leader-based top-down control ‘works’ when established cooperation dynamics are already in place, and the same can be said about distributed control (either as presented by a socially-positive cybernetics (e.g., Beer) or “free market”, invisible hand narratives). 

A starting point for the analysis of such multi-agent interactions may be from the point of view of game theory. In particular, the concept of Free-Energy Equilibria (FEE) captures the notion of a joint policy in which all agents in a given interaction are minimizing their (Expected) Free Energy \citep{hyland2024free}. Under certain conditions, it may be expected that interactions between agents tend to drive the system toward FEE. However, as is well understood in game theory, there are typically many equilibria of a game and depending on the starting point and interaction/learning dynamics, the system may converge to undesirable equilibria where a handful of agents exert significant control over others to their detriment. An important area of future study is thus to first establish a better understanding of what makes different equilibria more or less societally desirable, and then to understand how incentive landscapes may be shaped to navigate society toward better equilibria. Shared narratives, where agents are in the know about what is shared, are fundamental here, and this begins by paying attention to the (possible) rules in place, hence the need for formal methods. This is part of what critical metacognition entails: knowing how we are collectively navigating uncertainty, and taking account of already-given asymmetries. In the conclusion below, we explore a few salient implications of our arguments. 

\section{Conclusion}

Attentional power dynamics shape the probability landscape of future possibilities. Our proposal to address possibilistic power, grounded in AIF, offers ways to identify and address power imbalances. These formalizations can have significant impact for understanding power dynamics in an increasingly information-driven world. The transition from material to informational forms of power presents plenty of questions and opportunities for possible framings of future social dynamics. This paper examined how AIF can shed light on these shifts, suggesting that (possibilistic) power operates through mechanisms of social association, through the identification of and attention-pooling towards (apparently!) influential agents, as well as through enhanced computing-projective capabilities. Both these phenomena create self-reinforcing cycles, where empowered agents can expand their state space (and, often, minimize cost and future vulnerability). What has been previously framed in terms of territorial, material and cultural power can now also be understood in terms of informational and computational power, both of these have received attention as forms of capital; a term we can therefore reframe as possibilistic power. This new understanding links the cognitive foundations of narrative engagement and the ensuing social scripts which make or break power relations.

The ontoepistemic challenges in grounding (deontic) value can be understood under AIF in ways that open up complex semantics and social coordination towards a possible naturalization of these phenomena. Not in an effort to reduce them to simplistic images, rather in an effort to provide new entries into as well as possible accountability metrics of thorny ethical landscapes. In our proposal, coordinated behaviors emerge as shared models that reduce collective uncertainty (again: with the caveat that whatever is supposed to be “shared” is not always given nor understood). When scripts become internalized by multiple members of a community, this allows for the collective construction of future states. Following those who appear to have the potential to reduce our uncertainty, leads to agents associating with agents and actions that would seem to meet those expectations. Metacognitive criticality, that is: not just playing the game (or being played) but knowing the rules and being able to imagine new rules, is how we ought to collectively navigate uncertainty, assuming already given asymmetries. Future research could focus on developing metrics for measuring possibilistic power and designing interventions to promote equitable distribution of both material and informational resources.

This shift from territorial/material to informational/computational power requires special attention in futurological contexts. We propose that AIF provides crucial theoretical tools for making power structures visible and therefore analyzable, allowing for the potential to rethink current social asymmetries. Given the apparent transition between the organization of life around the centrality of physical work and matter, towards a centrality of attention and information, a sustainable social future requires addressing both resource and information distributions. The latter is best analyzed through the power of attention as a major factor influencing social dynamics.

Hierarchies are inevitable in complex thermodynamic systems. We propose fostering transparent, plastic hierarchies that remain contestable and therefore open to novelty. This approach acknowledges historic and thermodynamic constraints, while acknowledging space for systemic evolution. While hierarchies are inevitable, social scripts should be read-writable by those with an interest in survival within them. This helps reorient our understanding of an unfolding, globalizing, interconnected and interdependent life: participation in a system is only participatory if those involved are able to track the consequences of the variables and thus future state spaces at stake. A true distribution of information processing capabilities is essential for systemic sustainability and proper equitable development. And ``if nature is unjust: change nature.'' \citep{laboria2015xenofeminist}.

\bibliographystyle{alpha}
\bibliography{main}

\begin{thebibliography}{58}
\providecommand{\natexlab}[1]{#1}
\providecommand{\url}[1]{\texttt{#1}}
\expandafter\ifx\csname urlstyle\endcsname\relax
  \providecommand{\doi}[1]{doi: #1}\else
  \providecommand{\doi}{doi: \begingroup \urlstyle{rm}\Url}\fi

\bibitem[Albarracin et~al.(2021)Albarracin, Constant, Friston, and Ramstead]{albarracin2021variational}
Mahault Albarracin, Axel Constant, Karl~J Friston, and Maxwell James~D Ramstead.
\newblock A variational approach to scripts.
\newblock \emph{Frontiers in Psychology}, 12:\penalty0 585493, 2021.

\bibitem[Albarracin et~al.(2022)Albarracin, Demekas, Ramstead, and Heins]{albarracin2022epistemic}
Mahault Albarracin, Daphne Demekas, Maxwell~JD Ramstead, and Conor Heins.
\newblock Epistemic communities under active inference.
\newblock \emph{Entropy}, 24\penalty0 (4):\penalty0 476, 2022.

\bibitem[Andrews(2021)]{andrews2021math}
Mel Andrews.
\newblock The math is not the territory: navigating the free energy principle.
\newblock \emph{Biology \& Philosophy}, 36\penalty0 (3):\penalty0 30, 2021.

\bibitem[Barab{\'a}si and Albert(1999)]{barabasi1999emergence}
Albert-L{\'a}szl{\'o} Barab{\'a}si and R{\'e}ka Albert.
\newblock Emergence of scaling in random networks.
\newblock \emph{Science}, 286\penalty0 (5439):\penalty0 509--512, 1999.

\bibitem[Blakeley(2024)]{blakeley2024vulture}
Grace Blakeley.
\newblock \emph{Vulture Capitalism: Corporate Crimes, Backdoor Bailouts, and the Death of Freedom}.
\newblock Simon and Schuster, 2024.

\bibitem[Bouizegarene et~al.(2020)Bouizegarene, Ramstead, Constant, Friston, and Kirmayer]{bouizegarene2020narrative}
Nabil Bouizegarene, MJ~Ramstead, Axel Constant, Karl~J Friston, and Laurence~J Kirmayer.
\newblock Narrative as active inference: an integrative account of the functions of narratives.
\newblock \emph{Preprint}, 10\penalty0 (10.31234), 2020.

\bibitem[Bourdieu(2018)]{bourdieu2018structures}
Pierre Bourdieu.
\newblock Structures, habitus, practices.
\newblock In \emph{Rethinking the subject}, pages 31--45. Routledge, 2018.

\bibitem[Bruineberg(2017)]{bruineberg2017active}
Jelle Bruineberg.
\newblock Active inference and the primacy of the'i can'.
\newblock In \emph{Philosophy and predictive processing}, pages 1--18. MIND Group, 2017.

\bibitem[Cappelen et~al.(2024)Cappelen, Goldstein, and Hawthorne]{cappelen2024ai}
H~Cappelen, S~Goldstein, and J~Hawthorne.
\newblock Ai survival stories: a taxonomic analysis of ai existential risk.
\newblock preprint for Philosophy of AI Journal, 2024.

\bibitem[Constant et~al.(2019)Constant, Ramstead, Veissi{\`e}re, and Friston]{constant2019regimes}
Axel Constant, Maxwell~JD Ramstead, Samuel~PL Veissi{\`e}re, and Karl Friston.
\newblock Regimes of expectations: an active inference model of social conformity and human decision making.
\newblock \emph{Frontiers in psychology}, 10:\penalty0 679, 2019.

\bibitem[de~Jager(2021)]{dejager2021inevitably}
S~de~Jager.
\newblock Inevitably falling into place: Intuition as a function of spatial prediction.
\newblock In \emph{Philosophy after AI: Meaning and Understanding, Society for the Study of Artificial Intelligence and the Simulation of Behavior}, 2021.

\bibitem[de~Jager(2022)]{dejager2022fiction}
S~de~Jager.
\newblock Fiction, speculation, prediction.
\newblock preprint, 2022.

\bibitem[de~Jager(2023)]{dejager2023semantic}
S~de~Jager.
\newblock Semantic noise and conceptual stagnation in natural language processing.
\newblock \emph{Angelaki}, 28\penalty0 (3):\penalty0 111--132, 2023.

\bibitem[Deleuze(1992)]{deleuze1992postscript}
Gilles Deleuze.
\newblock Postscript on the societies of control.
\newblock \emph{October}, 59:\penalty0 3--7, 1992.

\bibitem[Dennett(1992)]{dennett1992self}
Daniel~C Dennett.
\newblock The self as a center of narrative gravity.
\newblock In F~Kessel, P~Cole, and D~Johnson, editors, \emph{Self and Consciousness: Multiple Perspectives}. Erlbaum, Hillsdale, NJ, 1992.

\bibitem[Fisher(2009)]{fisher2009capitalist}
Mark Fisher.
\newblock \emph{Capitalist realism: Is there no alternative?}
\newblock Zero Books, 2009.

\bibitem[Foucault(1982)]{foucault1982subject}
Michel Foucault.
\newblock The subject and power.
\newblock \emph{Critical Inquiry}, 8\penalty0 (4):\penalty0 777--795, 1982.

\bibitem[Foucault(1990)]{foucault1990history}
Michel Foucault.
\newblock \emph{The history of sexuality: An introduction, volume I}.
\newblock Vintage, New York, 1990.
\newblock Originally published 1978.

\bibitem[Friedman et~al.(2021)Friedman, Tschantz, Ramstead, Friston, and Constant]{friedman2021active}
Daniel~Ari Friedman, Alec Tschantz, Maxwell~JD Ramstead, Karl Friston, and Axel Constant.
\newblock Active inferants: An active inference framework for ant colony behavior.
\newblock \emph{Frontiers in behavioral neuroscience}, 15:\penalty0 647732, 2021.

\bibitem[Friston et~al.(2006)Friston, Kilner, and Harrison]{friston2006free}
Karl Friston, James Kilner, and Lee Harrison.
\newblock A free energy principle for the brain.
\newblock \emph{Journal of Physiology-Paris}, 100\penalty0 (1-3):\penalty0 70--87, 2006.

\bibitem[Friston et~al.(2015)Friston, Rigoli, Ognibene, Mathys, Fitzgerald, and Pezzulo]{friston2015active}
Karl Friston, Francesco Rigoli, Dimitri Ognibene, Christoph Mathys, Thomas Fitzgerald, and Giovanni Pezzulo.
\newblock Active inference and epistemic value.
\newblock \emph{Cognitive neuroscience}, 6\penalty0 (4):\penalty0 187--214, 2015.

\bibitem[Friston et~al.(2016)Friston, FitzGerald, Rigoli, Schwartenbeck, and Pezzulo]{friston2016active}
Karl Friston, Thomas FitzGerald, Francesco Rigoli, Philipp Schwartenbeck, and Giovanni Pezzulo.
\newblock Active inference: A process theory.
\newblock \emph{Neural Computation}, 29\penalty0 (1):\penalty0 1--49, 2016.
\newblock \doi{10.1162/NECO_a_00912}.

\bibitem[Friston et~al.(2010)Friston, Daunizeau, Kilner, and Kiebel]{friston2010action}
Karl~J Friston, Jean Daunizeau, James Kilner, and Stefan~J Kiebel.
\newblock Action and behavior: A free-energy formulation.
\newblock \emph{Biological Cybernetics}, 102:\penalty0 227--260, 2010.

\bibitem[Friston et~al.(2020)Friston, Parr, Yufik, Sajid, Price, and Holmes]{friston2020generative}
Karl~J Friston, Thomas Parr, Yan Yufik, Noor Sajid, Catherine~J Price, and Emma Holmes.
\newblock Generative models, linguistic communication and active inference.
\newblock \emph{Neuroscience \& Biobehavioral Reviews}, 118:\penalty0 42--64, 2020.

\bibitem[Friston et~al.(2024)Friston, Ramstead, Kiefer, Tschantz, Buckley, Albarracin, Pitliya, Heins, Klein, Millidge, et~al.]{friston2024designing}
Karl~J Friston, Maxwell~JD Ramstead, Alex~B Kiefer, Alexander Tschantz, Christopher~L Buckley, Mahault Albarracin, Riddhi~J Pitliya, Conor Heins, Brennan Klein, Beren Millidge, et~al.
\newblock Designing ecosystems of intelligence from first principles.
\newblock \emph{Collective Intelligence}, 3\penalty0 (1):\penalty0 26339137231222481, 2024.

\bibitem[Gigerenzer(2008)]{gigerenzer2008heuristics}
Gerd Gigerenzer.
\newblock Why heuristics work.
\newblock \emph{Perspectives on Psychological Science}, 3\penalty0 (1):\penalty0 20--29, 2008.

\bibitem[Golan and Foley(2022)]{golan2022understanding}
Amos Golan and Duncan~K Foley.
\newblock Understanding the constraints in maximum entropy methods for modeling and inference.
\newblock \emph{IEEE Transactions on Pattern Analysis and Machine Intelligence}, 45\penalty0 (3):\penalty0 3994--3998, 2022.

\bibitem[Hip{\'o}lito and van Es(2022)]{hipolito2022enactive}
In{\^e}s Hip{\'o}lito and Thomas van Es.
\newblock Enactive-dynamic social cognition and active inference.
\newblock \emph{Frontiers in Psychology}, 13:\penalty0 855074, 2022.

\bibitem[Hyland et~al.()Hyland, Gaven{\v{c}}iak, Da~Costa, Heins, Kovarik, Gutierrez, Wooldridge, and Kulveit]{hyland2024free}
David Hyland, Tom{\'a}{\v{s}} Gaven{\v{c}}iak, Lancelot Da~Costa, Conor Heins, Vojtech Kovarik, Julian Gutierrez, Michael~J Wooldridge, and Jan Kulveit.
\newblock Free-energy equilibria: Toward a theory of interactions between boundedly-rational agents.
\newblock In \emph{ICML 2024 Workshop on Models of Human Feedback for AI Alignment}.

\bibitem[Ispolatov et~al.(2012)Ispolatov, Ackermann, and Doebeli]{ispolatov2012division}
Iaroslav Ispolatov, Martin Ackermann, and Michael Doebeli.
\newblock Division of labour and the evolution of multicellularity.
\newblock \emph{Proceedings of the Royal Society B: Biological Sciences}, 279\penalty0 (1734):\penalty0 1768--1776, 2012.

\bibitem[Jaynes(1957)]{jaynes1957information}
Edwin~T Jaynes.
\newblock Information theory and statistical mechanics.
\newblock \emph{Physical Review}, 106\penalty0 (4):\penalty0 620--630, 1957.

\bibitem[Kastel et~al.(2023)Kastel, Hesp, Ridderinkhof, and Friston]{kastel2023small}
Natalie Kastel, Casper Hesp, K~Richard Ridderinkhof, and Karl~J Friston.
\newblock Small steps for mankind: Modeling the emergence of cumulative culture from joint active inference communication.
\newblock \emph{Frontiers in neurorobotics}, 16:\penalty0 944986, 2023.

\bibitem[Kent(2024)]{kent2024mental}
Lachlan Kent.
\newblock Mental gravity: Modelling the embodied self on the physical environment.
\newblock \emph{Journal of Environmental Psychology}, page 102245, 2024.

\bibitem[Kiverstein et~al.(2019)]{kiverstein2019obsessive}
Julian Kiverstein et~al.
\newblock Obsessive compulsive disorder: a pathology of self-confidence?
\newblock \emph{Trends in Cognitive Sciences}, 23\penalty0 (5):\penalty0 369--372, 2019.

\bibitem[Klyubin et~al.(2005)Klyubin, Polani, and Nehaniv]{klyubin2005empowerment}
Alexander~S Klyubin, Daniel Polani, and Chrystopher~L Nehaniv.
\newblock Empowerment: A universal agent-centric measure of control.
\newblock In \emph{Proceedings of the 2005 IEEE Congress on Evolutionary Computation}, pages 128--135, 2005.

\bibitem[Labella(2007)]{labella2007division}
Thomas~Halva Labella.
\newblock Division of labour in groups of robots.
\newblock \emph{Universite Libre de Bruxelles}, 2007.

\bibitem[{Laboria Cuboniks}(2015)]{laboria2015xenofeminist}
{Laboria Cuboniks}.
\newblock \emph{Xenofeminist Manifesto}.
\newblock 2015.

\bibitem[Lakoff and Johnson(2003)]{lakoff2003metaphors}
George Lakoff and Mark Johnson.
\newblock \emph{Metaphors we live by}.
\newblock University of Chicago Press, 2003.
\newblock Originally published 1980.

\bibitem[Leibfried et~al.(2019)Leibfried, Pascual-Diaz, and Grau-Moya]{leibfried2019unified}
Felix Leibfried, Sergio Pascual-Diaz, and Jordi Grau-Moya.
\newblock A unified bellman optimality principle combining reward maximization and empowerment.
\newblock \emph{Advances in Neural Information Processing Systems}, 32, 2019.

\bibitem[Maisto et~al.(2024)Maisto, Donnarumma, and Pezzulo]{maisto2024interactive}
Domenico Maisto, Francesco Donnarumma, and Giovanni Pezzulo.
\newblock Interactive inference: A multi-agent model of cooperative joint actions.
\newblock \emph{IEEE Transactions on Systems, Man, and Cybernetics: Systems}, 54\penalty0 (2):\penalty0 704--715, 2024.

\bibitem[Marzband et~al.(2017)Marzband, Ardeshiri, Moafi, and Uppal]{marzband2017distributed}
Mousa Marzband, Reza~Rouhi Ardeshiri, Milad Moafi, and Hassan Uppal.
\newblock Distributed generation for economic benefit maximization through coalition formation--based game theory concept.
\newblock \emph{International Transactions on Electrical Energy Systems}, 27\penalty0 (6):\penalty0 e2313, 2017.

\bibitem[Ororbia and Friston(2023)]{ororbia2023mortal}
Alexander Ororbia and Karl Friston.
\newblock Mortal computation: A foundation for biomimetic intelligence.
\newblock \emph{arXiv preprint arXiv:2311.09589}, 2023.

\bibitem[Parr et~al.(2019)Parr, Markovic, Kiebel, and Friston]{parr2019neuronal}
Thomas Parr, Dimitrije Markovic, Stefan~J Kiebel, and Karl~J Friston.
\newblock Neuronal message passing using mean-field, bethe, and marginal approximations.
\newblock \emph{Scientific Reports}, 9\penalty0 (1):\penalty0 1889, 2019.

\bibitem[Parr et~al.(2022)Parr, Pezzulo, and Friston]{parr2022active}
Thomas Parr, Giovanni Pezzulo, and Karl~J Friston.
\newblock \emph{Active Inference: The Free Energy Principle in Mind, Brain, and Behavior}.
\newblock MIT Press, Cambridge, MA, 2022.

\bibitem[Parvizi-Wayne(2024)]{parvizi2024preferences}
Darius Parvizi-Wayne.
\newblock How preferences enslave attention: calling into question the endogenous/exogenous dichotomy from an active inference perspective.
\newblock \emph{Phenomenology and the Cognitive Sciences}, pages 1--47, 2024.

\bibitem[Ram{\'i}rez-Ruiz et~al.(2024)Ram{\'i}rez-Ruiz, Grytskyy, Mastrogiuseppe, Habib, and Moreno-Bote]{ramirez2024complex}
Jorge Ram{\'i}rez-Ruiz, Dmytro Grytskyy, Chiara Mastrogiuseppe, Yasir Habib, and Ruben Moreno-Bote.
\newblock Complex behavior from intrinsic motivation to occupy future action-state path space.
\newblock \emph{Nature Communications}, 15\penalty0 (1), 2024.

\bibitem[Ramstead et~al.(2023)Ramstead, Sakthivadivel, Heins, Koudahl, Millidge, Da~Costa, Klein, and Friston]{ramstead2023bayesian}
Maxwell~JD Ramstead, Dalton~AR Sakthivadivel, Conor Heins, Magnus Koudahl, Beren Millidge, Lancelot Da~Costa, Brennan Klein, and Karl~J Friston.
\newblock On bayesian mechanics: a physics of and by beliefs.
\newblock \emph{Interface Focus}, 13\penalty0 (20220029), 2023.
\newblock \doi{10.1098/rsfs.2022.0029}.

\bibitem[Scott(2020)]{scott2020seeing}
James~C Scott.
\newblock \emph{Seeing like a state: How certain schemes to improve the human condition have failed}.
\newblock Yale University Press, 2020.

\bibitem[Shepard(1987)]{shepard1987toward}
Roger~N Shepard.
\newblock Toward a universal law of generalization for psychological science.
\newblock \emph{Science}, 237\penalty0 (4820):\penalty0 1317--1323, 1987.

\bibitem[Shteynberg et~al.(2023)]{shteynberg2023theory}
Garriy Shteynberg et~al.
\newblock Theory of collective mind.
\newblock \emph{Trends in Cognitive Sciences}, 27\penalty0 (11):\penalty0 1019--1031, 2023.

\bibitem[Simon(1978)]{simon1978rationality}
Herbert~A Simon.
\newblock Rationality as process and as product of thought.
\newblock \emph{The American Economic Review}, 68\penalty0 (2):\penalty0 1--16, 1978.

\bibitem[Tiomkin et~al.(2024)Tiomkin, Nemenman, Polani, and Tishby]{tiomkin2024intrinsic}
Stas Tiomkin, Ilya Nemenman, Daniel Polani, and Naftali Tishby.
\newblock Intrinsic motivation in dynamical control systems.
\newblock \emph{PRX Life}, 2\penalty0 (3):\penalty0 033009, 2024.

\bibitem[Tison and Poirier(2021)]{tison2021communication}
Remi Tison and Pierre Poirier.
\newblock Communication as socially extended active inference: An ecological approach to communicative behavior.
\newblock \emph{Ecological Psychology}, 33\penalty0 (3-4):\penalty0 197--235, 2021.

\bibitem[Vasil et~al.(2020)]{vasil2020world}
Jared Vasil et~al.
\newblock A world unto itself: human communication as active inference.
\newblock \emph{Frontiers in Psychology}, 11:\penalty0 417, 2020.

\bibitem[Volpi and Polani(2020)]{volpi2020goal}
Nicola~Catenacci Volpi and Daniel Polani.
\newblock Goal-directed empowerment: combining intrinsic motivation and task-oriented behavior.
\newblock \emph{IEEE Transactions on Cognitive and Developmental Systems}, 15\penalty0 (2):\penalty0 361--372, 2020.

\bibitem[Wareham et~al.(2023)]{wareham2023swarm}
Todd Wareham et~al.
\newblock Swarm control for distributed construction: A computational complexity perspective.
\newblock \emph{ACM Transactions on Human-Robot Interaction}, 12\penalty0 (1):\penalty0 1--45, 2023.

\bibitem[Watson and Scheidt(2005)]{watson2005autonomous}
David~P Watson and David~H Scheidt.
\newblock Autonomous systems.
\newblock \emph{Johns Hopkins APL Technical Digest}, 26\penalty0 (4):\penalty0 368--376, 2005.

\bibitem[Weber(2009)]{weber2009theory}
Max Weber.
\newblock \emph{The theory of social and economic organization}.
\newblock Simon and Schuster, 2009.

\end{thebibliography}
% \nocite{*}

\end{document}